\documentclass[letterpaper,journal]{IEEEtran}

\usepackage{amsmath,amssymb,amsfonts}
\usepackage{array}
\usepackage[caption=false,font=normalsize,labelfont=sf,textfont=sf]{subfig}
\usepackage{textcomp}
\usepackage{stfloats}
\usepackage{url}
\usepackage{verbatim}
\usepackage{graphicx}
\usepackage{cite}
\usepackage{tikz}

\usepackage{fancyhdr}
\fancypagestyle{arxivfirst}{%
  \fancyhf{}
  \fancyhead[C]{%
    \parbox{\textwidth}{%
      \normalfont\fontsize{8}{10}\selectfont\centering
      This work has been submitted to the IEEE for possible publication.\\
      Copyright may be transferred without notice, after which
      this version may no longer be accessible.\par
    }%
  }
  \fancyfoot[C]{\normalfont\footnotesize\thepage}

}

\graphicspath{{figures/}}
\ifdefined\pdfminorversion\pdfminorversion=7\fi

\begin{document}

\title{DC-Link Voltage Regulation in a Single-Isolation-Stage Multioutput Solid-State Transformer for Wide-Range DC Outputs}

\author{
Yi Hao,~\IEEEmembership{Student Member,~IEEE},
Levy Ferreira Costa,~\IEEEmembership{Senior Member,~IEEE},
Krzysztof Puczko,
Liangcai Shu,~\IEEEmembership{Member,~IEEE},
and Dongsheng Yang,~\IEEEmembership{Senior Member,~IEEE}%
\thanks{This work is supported by the European Union's Horizon Europe
research and innovation programme under the Marie Skłodowska-Curie
(MSCA) Doctoral Networks grant agreement No. 101072414 (E2GO).}%
\thanks{Y. Hao, L. F. Costa and D. Yang are with the Department of Electrical Engineering,
Eindhoven University of Technology, 5612 AP Eindhoven, The Netherlands
(e-mail: y.hao@tue.nl, l.costa@tue.nl; d.yang1@tue.nl).

K. Puczko is with Delta Electronics EMEA, 02-822 Warsaw, Poland
(e-mail: krzysztof.puczko@deltaww.com).

L. Shu is with Delta Electronics (Shanghai) Co., Ltd., Hangzhou, China
(e-mail: alan.shu@deltaww.com).}%
}

\markboth{}{}

\maketitle
\thispagestyle{arxivfirst}


\begin{abstract}
Single-isolation-stage solid-state transformers with balancing circuits eliminate an additional downstream dc--dc conversion stage. However, the remaining isolated converters must directly accommodate a wide output-voltage range. This paper investigates variable dc-link voltage operation of a single-isolation-stage multioutput SST to ease this requirement. Coordinated cascaded H-bridge modulation and embedded four-switch balancing circuits support intentionally unequal submodule dc-link voltages, allowing the isolated-converter input voltages to be adjusted according to the output requirements. The achievable voltage range is determined by relating the required balancing power to the BC transfer capability within the available CHB switching windows, subject to modulation and peak and rms current limits. The analysis quantifies the range extension provided by BC operation beyond modulation alone. For a single-phase-shift dual-active-bridge converter, input-voltage following within this range reduces the calculated peak current and extends the ideal zero-voltage-switching region compared with fixed input. Experiments on a six-submodule SST prototype demonstrate unequal dc-link voltage operation and inter-submodule power exchange. Experiments result on DAB demonstrate reduced current stress and improved efficiency with voltage following.
\end{abstract}


\section{Introduction}

\IEEEPARstart{T}{he}  increasing penetration of dc loads, renewable resources, energy-storage systems, and electric vehicles (EVs) is creating strong demand for power-conversion interfaces that combine medium-voltage connection, galvanic isolation, and flexible dc outputs. Solid-state transformers (SSTs) provide these functionalities through controllable power-electronic conversion and medium- or high-frequency galvanic isolation, and have therefore been extensively investigated for modern distribution systems and dc-load applications \cite{Huber2016SSTOrigins,She2013SSTReview,Saleh2019SSTPartI,Ruiz2020SSTSurvey}. Among the candidate medium-voltage front ends, the cascaded H-bridge (CHB) converter is particularly attractive because its series-connected submodules distribute the grid voltage across lower-voltage semiconductor devices while synthesizing a multilevel ac waveform. By associating each floating CHB dc link with an isolated dc--dc converter, a modular interface from the medium-voltage ac grid to multiple low-voltage dc loads can be obtained \cite{Zhao2013VoltagePowerBalance,Wang2016PowerVoltageBalance}.
A single-isolation-stage CHB-based SST architecture was previously proposed in \cite{Hao2026SingleIsolation} for multioutlet EV fast-charging stations, in which the load is connected directly to the isolated dc--dc converter associated with each CHB submodule and embedded balancing circuits (BCs) redistribute power among submodules to accommodate unequal port loads. Eliminating an additional downstream dc--dc stage reduces the number of cascaded conversion stages, but it also requires the remaining isolated converter to accommodate the load-voltage range directly. The design of the isolated dc--dc stage is already one of the main challenges in modular SSTs because its topology, semiconductor stress, magnetic design, efficiency, and operating range directly affect the overall system performance \cite{Franca2026DCDCReview}. Accommodating a wide output-voltage range makes this design
particularly challenging, with recently reported converter designs targeting as wide as 200--1000~V \cite{Aarninkhof2024WideRange,Elezab2023WideRangeLLC,Zayed2023WideRangeDAB}. With a fixed isolated-converter input voltage, such a wide output span requires a correspondingly wide conversion-ratio range. Wide-gain LLC resonant converters and dual-active-bridge (DAB) converters therefore commonly introduce additional operating modes, reconfigurable structures, or more sophisticated modulation in order to retain favorable soft-switching, current-stress, and efficiency characteristics over the full voltage range \cite{Zhao2023HybridLLC,Xie2024VTDAB,Zhao2014DABOverview}. These observations motivate using the CHB submodule dc-link voltage itself as an additional voltage-design degree of freedom.

Conventional modular CHB-based SST control, however, generally treats the individual floating dc-link voltages as quantities to be balanced to common values. Representative voltage- and power-balancing methods regulate the cascaded cell voltages while coordinating the power processed by the isolated stages \cite{Zhao2013VoltagePowerBalance,Wang2016PowerVoltageBalance}. Decentralized droop-based strategies similarly restore dc-link voltage and power balance among modules \cite{Rodrigues2018AdaptiveDroop}, while naturally balanced resonant stages have been proposed to enforce equal cell power and voltage sharing through the dc--dc stage itself \cite{Bhawal2023NaturalBalance}. More recent work extends dc-link balancing toward wider loading conditions \cite{Burugula2024DCLinkBalancing}, and hardware-assisted submodule-parallelization methods directly equalize adjacent CHB capacitor voltages \cite{Zhang2024SensorlessCHB}. A related body of work addresses unequal module powers through CHB power-routing methods, for example by modifying the modulation indices or injecting harmonic components to enlarge the allowable power-disparity range \cite{Ko2017PowerRouting,Ibrahim2026PowerRouting}. Although these approaches improve power-sharing flexibility, their objective is to manage unequal power while maintaining or restoring the internal dc-link voltage balance, rather than to intentionally assign different dc-link operating voltages according to the conversion-ratio requirements of the downstream isolated converters.

Beyond internal voltage and power balancing, research on multiport SST and power electronic transformer (PET) architectures has expanded the ability to interface diverse sources and loads. Existing studies have demonstrated coordinated operation of multiple ac and dc ports through integrated energy storage, power-decoupling structures, and partial-power-processing control \cite{Vasiladiotis2015MultiportPET,Farooqi2024MultiportPET,Ma2020MultiportPET,Zheng2022ThreePortSST,Zheng2023MultiportPPP}.
However, flexibility at the external ports does not necessarily imply that internal submodule dc-link voltages are used as adjustable variables to improve the operation of the isolated conversion stage. For example, extensible-port SSTs can provide independent dc buses with different voltage levels and capacities, while internal submodule capacitor-voltage deviations are still suppressed through coordinated power redistribution \cite{Xia2026ExtensiblePortSST}. These developments motivate further investigation of intentionally unequal submodule dc-link voltages and their
potential to ease the conversion-ratio requirements of the downstream isolated dc--dc converters.

This paper investigates intentionally unequal CHB dc-link voltages as operating variables for adapting the inputs of the isolated dc--dc converters to their output requirements. The selected voltage distribution is supported by coordinated CHB modulation and embedded BCs between adjacent submodules. Operating as four-switch buck--boost converters within the available CHB switching windows, the BCs transfer the differential power arising from the selected voltage and load distributions.

The central question is how far the submodule voltages can be adjusted within the modulation and BC current limits, and how this voltage flexibility benefits the isolated conversion stage. This paper addresses these questions through a local submodule-pair model, DAB analysis, and experimental evaluation. 

The main contributions of this paper are summarized as follows:
\begin{itemize}

\item Variable dc-link voltage operation is developed for a single-isolation-stage multioutput CHB-SST using coordinated CHB modulation and embedded four-switch BCs. The available CHB switching windows and the corresponding BC operating states are identified.

\item A voltage-regulation model is established for an adjacent submodule pair. It relates the required balancing power to the BC transfer capability under modulation and current constraints, and determines the voltage-range extension beyond modulation-only operation.

\item The derived voltage range is applied to a DAB converter to compare fixed input with voltage following under identical hardware and output requirements. Peak-current reduction and extension of the ideal ZVS region are quantified.

\item A six-submodule SST prototype demonstrates unequal dc-link voltage operation and BC power exchange. Separate DAB tests evaluate the resulting current-stress and efficiency benefits of input-voltage adjustment.

\end{itemize}

The remainder of this paper is organized as follows. Section~\ref{sec:operating_principle} introduces the system configuration, BC switching states, and modulation. Section~\ref{sec:voltage_regulation_range} establishes the current-constrained voltage-regulation region. Section~\ref{sec:dab_benefits} quantifies the peak-current and ideal ZVS benefits of BC-constrained input-voltage following for the isolated dc--dc stage. Section~\ref{sec:experiment} presents the experimental results, and Section~\ref{sec:conclusion} concludes the paper.

\section{System Configuration and Operating Principle}
\label{sec:operating_principle}

\subsection{System Architecture}
\label{subsec:system_architecture}

\begin{figure}[!t]
\centering
\includegraphics[width=\columnwidth]{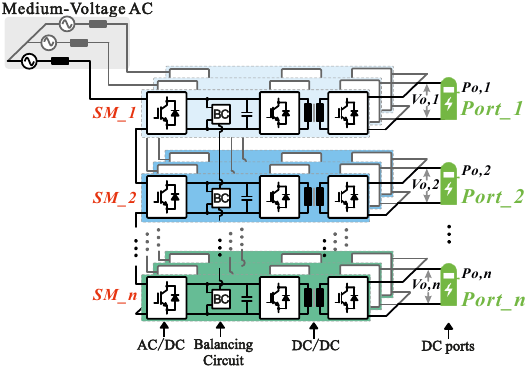}
\caption{Single-isolation-stage multioutput SST with embedded balancing circuits. The output-port voltages $V_{o,k}$ are distinct from the submodule dc-link voltages $V_k$.}
\label{fig:sst_architecture}
\end{figure}

Fig.~\ref{fig:sst_architecture} shows the single-isolation-stage multioutput SST. The grid-side CHB converter supplies the floating submodule dc links, and the associated isolated dc--dc converters directly feed the output ports. Embedded balancing circuits provide additional power-transfer paths between adjacent submodules. By redistributing differential power, these circuits support intentionally unequal dc-link voltages, allowing the isolated-converter input voltages to be adjusted according to the output requirements.

\begin{figure}[!t]
\centering
\includegraphics[width=0.88\columnwidth]{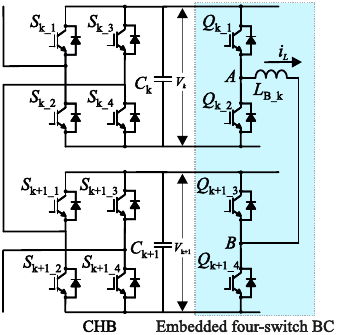}
\caption{Topology of the CHB with embedded four-switch BC }
\label{fig:embedded_bc}
\end{figure}

Fig.~\ref{fig:embedded_bc} shows the topology of the embedded four-switch BC between adjacent CHB submodules $k$ and $k+1$. The capacitors $C_k$ and $C_{k+1}$ support the dc-link voltages $V_k$ and $V_{k+1}$, respectively. The highlighted BC comprises two switching legs connected across the corresponding dc links: $Q_{k,1}$--$Q_{k,2}$ and $Q_{k+1,3}$--$Q_{k+1,4}$. The balancing inductor $L_{B,k}$ connects their midpoints $A$ and $B$. The inductor current $i_L$ is defined as positive from $A$ to $B$, giving $v_L=v_A-v_B=L_{B,k}\,di_L/dt$. Forward balancing power denotes net transfer from module $k$ to module $k+1$.

The adjacent CHB submodules share an ac-side node, while their dc links remain floating. The CHB switching states therefore determine the instantaneous connection between the dc links and the available BC operating windows, as described in the following subsection.

\subsection{BC Operating Windows and Switching States}
\label{subsec:operating_windows}
\label{subsec:buck_boost_states}

The BC operates within switching intervals established by the adjacent CHB submodules. As shown in Fig.~\ref{fig:embedded_bc}, the upper-module right CHB leg and the lower-module left CHB leg share an ac-side node. When $S_{k,4}$ and $S_{k+1,2}$ are on, this node connects the negative dc-link terminals, establishing the common-negative window $W_{\mathrm N}$. When $S_{k,3}$ and $S_{k+1,1}$ are on, the positive terminals are connected instead, establishing the common-positive window $W_{\mathrm P}$. These connections provide the return paths required for the selected BC power-transfer sequences.

Forward transfer from module $k$ to module $k+1$ uses positive inductor current in $W_{\mathrm N}$ and negative current in $W_{\mathrm P}$. This polarity change results from the different common-terminal connections and does not reverse the balancing-power direction. In the following state descriptions, the common terminal is taken as the local zero-potential reference. The fixed current arrow denotes the reference direction from $A$ to $B$, whereas the dashed arrows indicate the actual BC-loop current direction.

For forward transfer, buck operation applies when $V_k>V_{k+1}$ and boost operation when $V_k<V_{k+1}$. Each mode uses two switching states within either window. The descriptions below concern the BC-loop current; the shared CHB switches may additionally carry the grid current.

\begin{figure}[!t]
\centering
\begin{minipage}[b]{0.48\columnwidth}
\centering
\subfloat[Buck mode: N-State I]{\includegraphics[width=\linewidth]{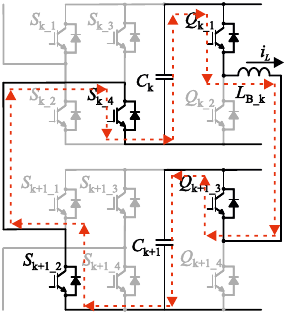}\label{fig:bc_buck_1}}
\end{minipage}
\hfill
\begin{minipage}[b]{0.48\columnwidth}
\centering
\subfloat[Buck mode: N-State II]{\includegraphics[width=\linewidth]{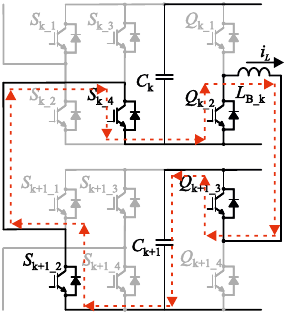}\label{fig:bc_buck_2}}
\end{minipage}

\vspace{2mm}
\begin{minipage}[b]{0.48\columnwidth}
\centering
\subfloat[Buck mode: P-State I]{\includegraphics[width=\linewidth]{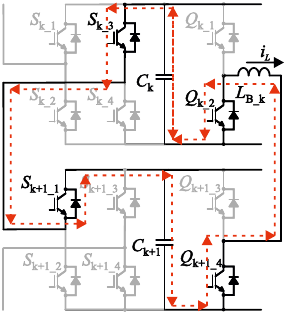}\label{fig:bc_buck_3}}
\end{minipage}
\hfill
\begin{minipage}[b]{0.48\columnwidth}
\centering
\subfloat[Buck mode: P-State II]{\includegraphics[width=\linewidth]{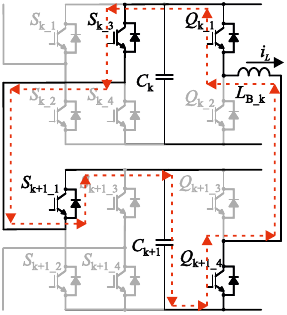}\label{fig:bc_buck_4}}
\end{minipage}
\caption{BC switching states for forward buck transfer: (a) N-State I, (b) N-State II, (c) P-State I, and (d) P-State II. Dashed arrows indicate the BC-loop current paths.}
\label{fig:bc_buck_states}
\end{figure}

In buck operation, Fig.~\ref{fig:bc_buck_states} shows the two states for each window. In $W_{\mathrm N}$, $Q_{k+1,3}$ remains on while $Q_{k,1}$ and $Q_{k,2}$ switch complementarily. N-State I connects $A$ and $B$ to the positive rails through $Q_{k,1}$ and $Q_{k+1,3}$, giving $v_L=V_k-V_{k+1}>0$. The source supplies the receiving module while the inductor current increases. In N-State II, $Q_{k,2}$ connects $A$ to the common negative rail, giving $v_L=-V_{k+1}$. The inductor continues to supply module $k+1$ while its positive current decreases, with no current drawn from $C_k$ through the BC.

In the buck P window, $Q_{k+1,4}$ remains on and the upper-leg switching roles are exchanged. P-State I turns on $Q_{k,2}$, placing the midpoints at $v_A=-V_k$ and $v_B=-V_{k+1}$. Consequently, $v_L=V_{k+1}-V_k<0$, and the negative current increases in magnitude. P-State II turns on $Q_{k,1}$ instead, giving $v_L=V_{k+1}>0$. The negative-current magnitude then decreases as the inductor continues supplying the receiving module. Thus, in both N and P windows, buck State I increases the forward-current magnitude and State II decreases it.

\begin{figure}[!t]
\centering
\begin{minipage}[b]{0.48\columnwidth}
\centering
\subfloat[Boost mode: N-State I]{\includegraphics[width=\linewidth]{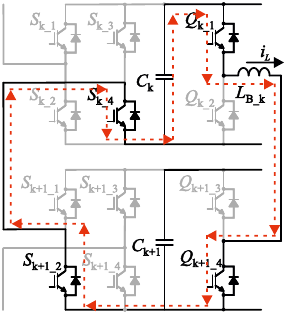}\label{fig:bc_boost_1}}
\end{minipage}
\hfill
\begin{minipage}[b]{0.48\columnwidth}
\centering
\subfloat[Boost mode: N-State II]{\includegraphics[width=\linewidth]{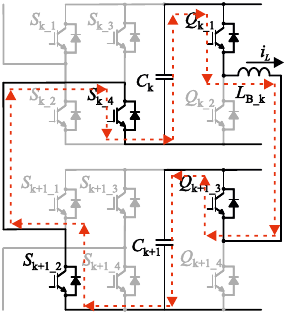}\label{fig:bc_boost_2}}
\end{minipage}

\vspace{2mm}
\begin{minipage}[b]{0.48\columnwidth}
\centering
\subfloat[Boost mode: P-State I]{\includegraphics[width=\linewidth]{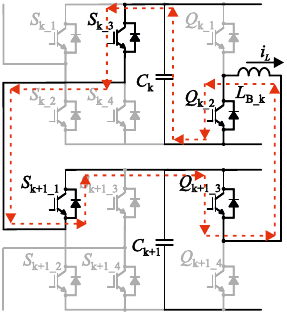}\label{fig:bc_boost_3}}
\end{minipage}
\hfill
\begin{minipage}[b]{0.48\columnwidth}
\centering
\subfloat[Boost mode: P-State II]{\includegraphics[width=\linewidth]{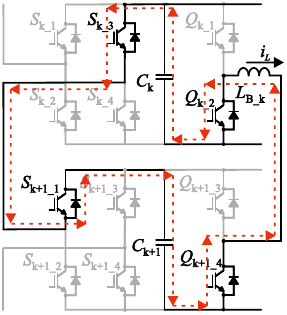}\label{fig:bc_boost_4}}
\end{minipage}
\caption{BC switching states for forward boost transfer: (a) N-State I, (b) N-State II, (c) P-State I, and (d) P-State II. Dashed arrows indicate the BC-loop current paths.}
\label{fig:bc_boost_states}
\end{figure}

In boost operation, the states are shown in Fig.~\ref{fig:bc_boost_states}. In $W_{\mathrm N}$, $Q_{k,1}$ remains on while the lower BC leg switches complementarily. N-State I turns on $Q_{k+1,4}$, applying $v_L=V_k$ and storing energy in the inductor without delivering BC current to $C_{k+1}$. N-State II turns on $Q_{k+1,3}$ instead, giving $v_L=V_k-V_{k+1}<0$. The source and inductor supply module $k+1$, and the positive current decreases.

In the boost P window, $Q_{k,2}$ remains on. P-State I turns on $Q_{k+1,3}$, giving $v_L=-V_k$ and increasing the negative-current magnitude as energy is stored. P-State II turns on $Q_{k+1,4}$ instead, giving $v_L=V_{k+1}-V_k>0$. The negative-current magnitude decreases while energy is delivered to module $k+1$. The N and P sequences therefore implement the same charging and transfer functions with opposite current polarities.

Let $J=|i_L|$ during intended forward transfer and let $d$ denote the fraction of the BC switching period occupied by State I. The N and P sequences yield the same switching-period-averaged magnitude dynamics:
\begin{equation}
L_{B,k}\frac{dJ}{dt}=
\begin{cases}
dV_k-V_{k+1},&\text{buck},\\
V_k-(1-d)V_{k+1},&\text{boost}.
\end{cases}
\label{eq:op_average}
\end{equation}
Constant average current requires $d=V_{k+1}/V_k$ for buck transfer and $d=1-V_k/V_{k+1}$ for boost transfer. These duty ratios describe regulated transfer after the current has been established. Startup and residual-current reset are included in the capability analysis of Section~\ref{sec:voltage_regulation_range}.

\subsection{BC Modulation Sequence}
\label{subsec:bc_modulation}

\begin{figure}[!t]
\centering
\includegraphics[width=\columnwidth]{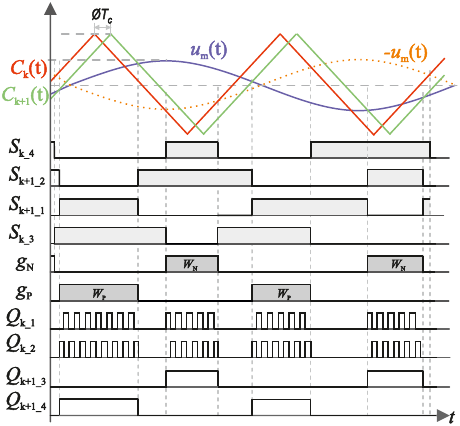}
\caption{Coordination of CHB modulation and basic buck-mode BC gate signals: carrier comparison, N/P window detection, and BC switching within the available windows. Window-entry current transients are omitted, and the switching-frequency ratio is not drawn to scale.}
\label{fig:bc_modulation}
\end{figure}

The switching states are coordinated with phase-shifted-carrier unipolar SPWM of the CHB. Fig.~\ref{fig:bc_modulation} illustrates this coordination for buck operation, using identical sinusoidal modulation references and displaced carriers for the adjacent submodules. The CHB gate signals first determine the available windows through $g_{\mathrm N}=S_{k,4}\land S_{k+1,2}$ and $g_{\mathrm P}=S_{k,3}\land S_{k+1,1}$, where $\land$ denotes logical AND. These indicators enable the corresponding BC switching sequence.

Within $W_{\mathrm N}$, the upper BC leg alternates between buck N-States I and II, while the lower leg maintains the connection through $Q_{k+1,3}$. Within $W_{\mathrm P}$, the connection changes to $Q_{k+1,4}$ and the upper-leg duty assignments exchange: $Q_{k,1}$ realizes State I in the N window, whereas $Q_{k,2}$ realizes State I in the P window. Consequently, the State-I duty ratio $d$ produces the same average current-magnitude dynamics in both windows. Boost uses the same window indicators but applies high-frequency switching to the lower BC leg according to Fig.~\ref{fig:bc_boost_states}.

Outside the available windows, all BC gate commands are disabled. Residual current may continue through the diode paths, so active-window duration and current-conduction duration are not identical. The finite window lengths, together with current buildup and subsequent decay, determine the net power that the BC can transfer under its current limits.

\section{Achievable Submodule Voltage-Regulation Range}
\label{sec:voltage_regulation_range}

The submodule dc-link voltages can be redistributed by adjusting the CHB modulation amplitudes and transferring differential power through the BC. This section first determines the voltage range obtainable by modulation alone and then evaluates its extension with bidirectional BC transfer. The modulation references retain the same sinusoidal phase and frequency, while their amplitudes are independently adjustable within the linear modulation range. The BC switching period is assumed much shorter than the CHB carrier period, and the dc-link voltages and modulation references are treated as locally constant when calculating the current trajectories.

\subsection{Power Allocation and Balancing Requirements}
\label{subsec:vr_power_balance}
The analysis considers two adjacent submodules within the SST. Voltage redistribution and BC power exchange are confined to this pair, while the remaining submodules retain their nominal dc-link voltages and modulation settings. The pair maintains its total dc-link voltage and fundamental ac-voltage contribution as its internal voltage allocation changes.

Let $V_{\mathrm{sm}}$ and $P_{\mathrm{sm}}$ denote the nominal submodule voltage and rated output power, respectively, and let $M_0=0.85$ denote the nominal modulation amplitude. The modulation amplitudes of the two selected submodules are denoted by $M_1$ and $M_2$. With $r=V_1/V_2>0$ and $v_i=V_i/V_{\mathrm{sm}}$, the pair constraints are
\begin{equation}
\begin{aligned}
v_1&=\frac{2r}{1+r},\qquad v_2=\frac{2}{1+r},\\
M_1v_1+M_2v_2&=2M_0,\qquad 0\leq M_1,M_2\leq1.
\end{aligned}
\label{eq:vr_voltages}
\end{equation}

The two submodules carry the same ac current. With sinusoidal modulation references of the same phase, their average ac-side powers are proportional to $M_1V_1$ and $M_2V_2$. Let $x=P_{\mathrm{ac},1}/(P_{\mathrm{ac},1}+P_{\mathrm{ac},2})$ denote the fraction of the pair's ac input power allocated to module~1. The corresponding modulation amplitudes are
\begin{equation}
x=\frac{M_1v_1}{2M_0},\qquad M_1=\frac{2M_0x}{v_1},\qquad M_2=\frac{2M_0(1-x)}{v_2}.
\label{eq:vr_modulation_allocation}
\end{equation}
For a given voltage ratio $r$, specifying $x$ determines both modulation amplitudes. Their limits require $x_{\mathrm{min}}\leq x\leq x_{\mathrm{max}}$, where

\begin{equation}
\begin{aligned}
x_{\mathrm{min}}&=\max\left\{0,1-\frac{v_2}{2M_0}\right\},\\
x_{\mathrm{max}}&=\min\left\{1,\frac{v_1}{2M_0}\right\}.
\end{aligned}
\label{eq:vr_allocation_limits}
\end{equation}

Let $\lambda=P_1/P_2$ denote the load-power ratio and $\alpha=P_2/P_{\mathrm{sm}}$ the loading level of module~2, where $0<\alpha\leq1$ and $0\leq\alpha\lambda\leq1$. Neglecting losses, the pair receives $P_1+P_2$ from the ac side. Subtracting the load of module~1 from its allocated ac-side power gives the signed balancing-power requirement:
\begin{equation}
p_{\mathrm{req}}\equiv\frac{P_{\mathrm{BC,req}}}{P_{\mathrm{sm}}}=\alpha\big[(1+\lambda)x-\lambda\big].
\label{eq:vr_required_power}
\end{equation}
Positive $p_{\mathrm{req}}$ denotes transfer from module~1 to module~2, whereas a negative value denotes transfer in the opposite direction. Equal modulation amplitudes give $x=r/(1+r)$ and recover $p_{\mathrm{req}}=\alpha(r-\lambda)/(1+r)$. 
With the BC disabled, each submodule supplies its own load, so $x=\lambda/(1+\lambda)$. Substituting this allocation into \eqref{eq:vr_allocation_limits} gives the modulation-only voltage range:
\begin{equation}
\frac{M_0\lambda}{1+\lambda-M_0\lambda}\leq r\leq\frac{1+\lambda-M_0}{M_0},\qquad r>0.
\label{eq:vr_modulation_only_range}
\end{equation}
Thus, modulation alone can support unequal dc-link voltages when the required ac-side power allocation remains within the modulation limits. Enabling the BC permits a different allocation, provided that the resulting differential power can be transferred within its operating windows and current limits.

\subsection{BC Operating Windows and Voltage-Regulation Limits}
\label{subsec:vr_transfer_model}
\label{subsec:vr_current_references}
\label{subsec:vr_feasible_region}

Within each carrier period, the fundamental phase $\theta$ is treated as constant, giving the modulation references $\mu_1=M_1\sin\theta$ and $\mu_2=M_2\sin\theta$.
The triangular carriers have period $T_c$ and relative time displacement $\phi T_c$. The switching states of the right CHB leg of module~1 and the left CHB leg of module~2 determine the available BC windows.
Neglecting dead time, $S_{1,3}=S_{2,1}=0$ connects the negative dc-link terminals and forms $W_{\mathrm N}$, while $S_{1,3}=S_{2,1}=1$ connects the positive terminals and forms $W_{\mathrm P}$.
The window indicators and their total durations per carrier period are

\begin{equation}
\begin{aligned}
g_{\mathrm P}&=S_{1,3}\land S_{2,1},\\
g_{\mathrm N}&=(\neg S_{1,3})\land(\neg S_{2,1}),\\
T_w(\theta)&=\int_0^{T_c}g_w(t;\theta)\,dt,
\qquad w\in\{\mathrm N,\mathrm P\}.
\end{aligned}
\label{eq:vr_window}
\end{equation}

Here, each switching variable is one for an on-state command and zero otherwise; $\land$ and $\neg$ denote logical AND and NOT, respectively. The window indicators likewise take values of zero or one.

For unequal modulation amplitudes, the window positions and durations are obtained directly from the carrier comparisons. The current calculation retains the complete switching-state sequence, including the intervals between active windows, to account for residual-current conduction.
The following model describes power transfer from module~1
to module~2. Let $\bar i_L$ denote the BC switching-period-averaged
current and $J=|\bar i_L|$ its absolute value. During active
forward transfer, $\bar i_L=J$ in $W_{\mathrm N}$ and
$\bar i_L=-J$ in $W_{\mathrm P}$.

Neglecting switching ripple, the current dynamics and port powers are
\begin{equation}
\begin{aligned}
L_B\frac{dJ}{dt}&=aV_1-bV_2,\\
\bar p_{\mathrm B,1}&=aV_1J,\qquad
\bar p_{\mathrm B,2}=bV_2J.
\end{aligned}
\label{eq:vr_active_model}
\end{equation}
Here, $a$ and $b$ describe the effective connection of the source and receiving dc links, respectively, as determined by the BC switching states and duty ratio. The port powers $\bar p_{\mathrm B,1}$ and $\bar p_{\mathrm B,2}$ are the averaged powers extracted from module~1 and delivered to module~2. The current builds toward a constant reference $J^*$ using $(a,b)=(1,0)$ and is then maintained using $(a,b)=(1,r)$ for boost or $(1/r,1)$ for buck operation. At $r=1$, both reduce to $(a,b)=(1,1)$. All BC gates are turned off at the window boundary, even if $J^*$ has not been reached.

With the BC disabled, nonzero residual current follows the diode paths. Using the signed current, their connection coefficients are
\begin{equation}
(a_o,b_o)=
\begin{cases}
(-S_{1,3},1-S_{2,1}),&\bar i_L>0,\\
(1-S_{1,3},-S_{2,1}),&\bar i_L<0.
\end{cases}
\label{eq:vr_off_coefficients}
\end{equation}
The off-state dynamics are $L_B\,d\bar i_L/dt=a_oV_1-b_oV_2$, with port powers $\bar p_{\mathrm B,1}=a_oV_1\bar i_L$ and $\bar p_{\mathrm B,2}=b_oV_2\bar i_L$. Negative input-port power represents energy returned to module~1. When both coefficients vanish, the current freewheels without exchanging capacitor energy but still contributes to rms stress. Diode conduction ceases when the current reaches zero.

If residual current opposes the intended transfer direction at a window entrance, the BC gates remain off until the current reaches zero. Active transfer begins only if time remains in the window; otherwise, residual-current conduction continues according to \eqref{eq:vr_off_coefficients}.

For each fixed fundamental phase $\theta$, the current trajectory is calculated over successive carrier periods until periodic steady state is reached. The resulting trajectories are used to calculate the net transferred power, rms current, and peak current. With $\langle\cdot\rangle$ denoting averaging over the carrier period and fundamental phase, these quantities are

\begin{equation}
\begin{aligned}
P_{\mathrm{BC}}&=\langle\bar p_{\mathrm B,2}\rangle=\langle\bar p_{\mathrm B,1}\rangle,\\
I_{L,\mathrm{rms}}&\simeq\sqrt{\langle\bar i_L^2\rangle},\qquad
I_{L,\mathrm{pk}}\simeq\max_{\theta,t}|\bar i_L|.
\end{aligned}
\label{eq:vr_phase_average}
\end{equation}

Let $I_b=P_{\mathrm{sm}}/V_{\mathrm{sm}}$ be the current base and $\ell=L_BP_{\mathrm{sm}}/(V_{\mathrm{sm}}^2T_c)$ the normalized inductance. At the nominal unity-power-factor operating point, the rated CHB grid-current peak and rms values are $2I_b/M_0$ and $\sqrt{2}I_b/M_0$, respectively. For a consistent normalized comparison, these values are adopted as reference limits for the BC inductor current:
\begin{equation}
\frac{I_{L,\mathrm{pk}}}{I_b}\leq\frac{2}{M_0},\qquad
\frac{I_{L,\mathrm{rms}}}{I_b}\leq\frac{\sqrt{2}}{M_0}.
\label{eq:vr_normalized_limits}
\end{equation}
The normalized limits are approximately $2.353$ and $1.664$. They apply to the BC inductor current and remain fixed as the modulation amplitudes and loads change.

For each $(r,x)$, the current reference $J^*$ is increased from zero until the rms-current limit is reached or $J^*=2I_b/M_0$, whichever occurs first. Under the adopted averaged-current policy, the current magnitude does not exceed $J^*$, so the reference upper limit also enforces the peak-current constraint. The transferred power at the selected reference defines $p_{\mathrm{cap}}(r,x;\ell)=P_{\mathrm{BC}}/P_{\mathrm{sm}}$ under this policy. As with $p_{\mathrm{req}}$, the power is normalized by $P_{\mathrm{sm}}$. Assuming a continuous reference-to-power characteristic, intermediate positive powers are obtained by selecting a reference within the admissible interval.

A voltage ratio is feasible under joint modulation and bidirectional BC transfer if an admissible allocation exists such that

\begin{equation}
\exists\,x\in[x_{\mathrm{min}},x_{\mathrm{max}}]:\quad
\left|\alpha\big[(1+\lambda)x-\lambda\big]\right|
\leq p_{\mathrm{cap}}^{\mathrm{dir}}(r,x;\ell).
\label{eq:vr_power_feasibility}
\end{equation}

where $p_{\mathrm{cap}}^{\mathrm{dir}}$ uses module~1 as the source when $p_{\mathrm{req}}\geq0$ and module~2 as the source when $p_{\mathrm{req}}<0$. The reverse capability is evaluated by exchanging $(V_1,M_1)$ and $(V_2,M_2)$ in the same trajectory model. The search therefore accounts for both effects of modulation allocation: it changes the magnitude and direction of the required balancing power as well as the available BC windows. For each $r$, the allocation is varied to find the largest difference between capability and demand. The zero-power allocation is included with the BC disabled, so the joint feasible region contains the modulation-only range. The endpoints of each connected feasible interval define the voltage-ratio bounds under this control policy.

\subsection{Achievable Voltage Range}
\label{subsec:vr_results}

\begin{figure*}[!t]
\centering
\begin{tikzpicture}
\node[inner sep=0pt] (main) {%
\begin{minipage}{\textwidth}
\centering

\subfloat[]{%
    \includegraphics[
        width=0.52\textwidth,
        trim={0mm 12mm 0mm 15mm},
        clip
    ]{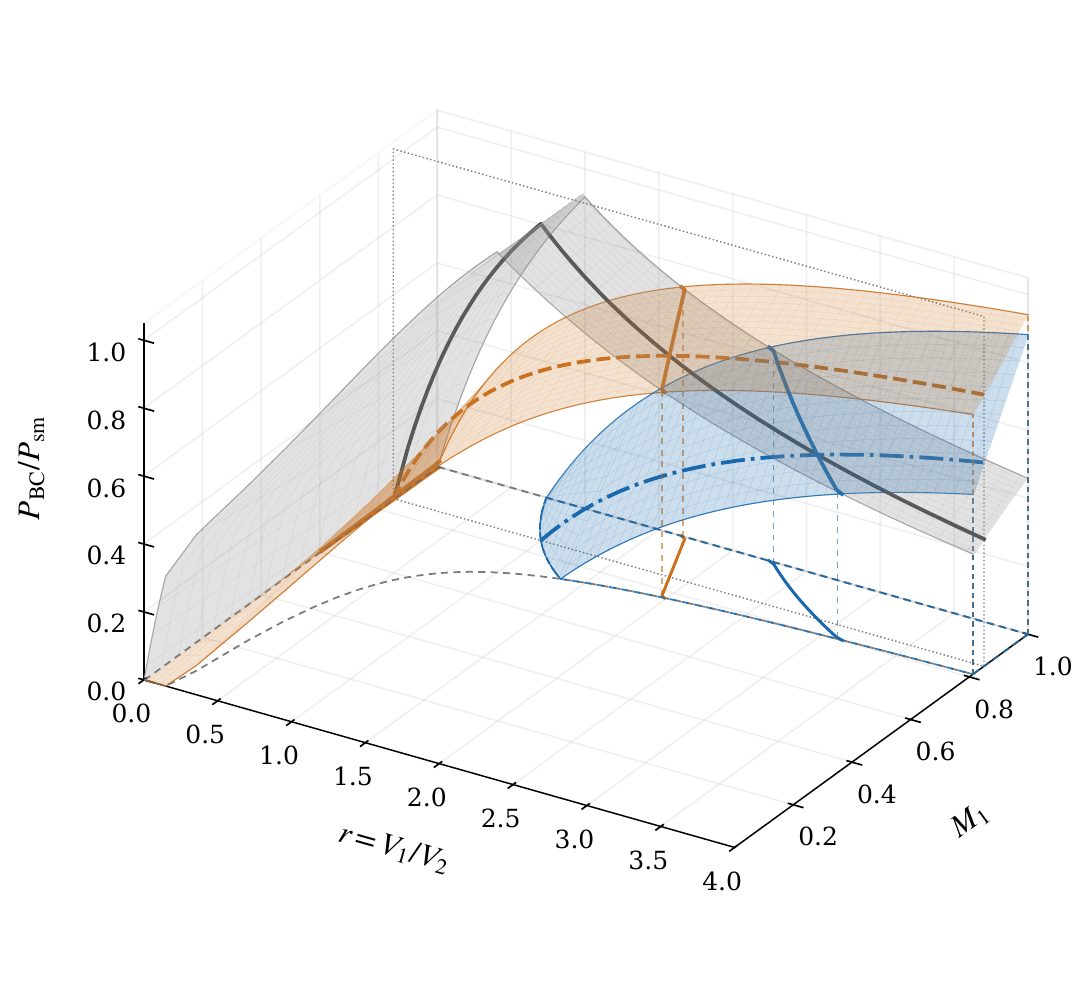}%
    \label{fig:bc_surfaces}}
\hfill
\subfloat[]{%
    \includegraphics[
        width=0.44\textwidth,
        trim={0mm 4mm 0mm 10mm},
        clip
    ]{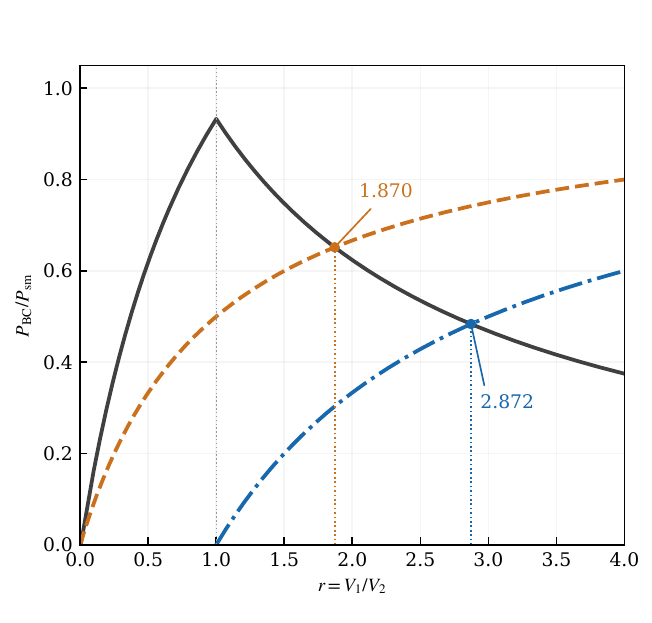}%
    \label{fig:bc_power_section}}

\par\vspace{-3mm}

\subfloat[]{%
    \includegraphics[
        width=0.44\textwidth,
        trim={0mm 4mm 0mm 10mm},
        clip
    ]{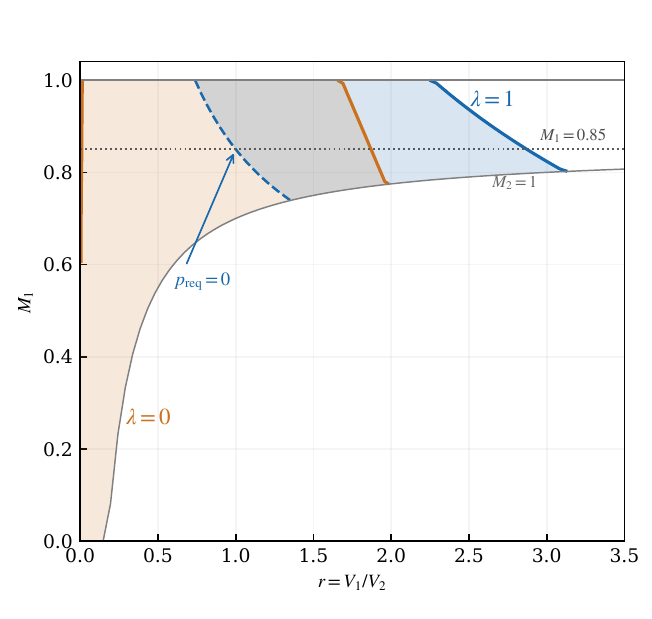}%
    \label{fig:bc_modulation_projection}}
\hfill
\subfloat[]{%
    \raisebox{0mm}{%
    \includegraphics[
        width=0.44\textwidth,
        trim={0mm 2mm 0mm 7mm},
        clip
        ]{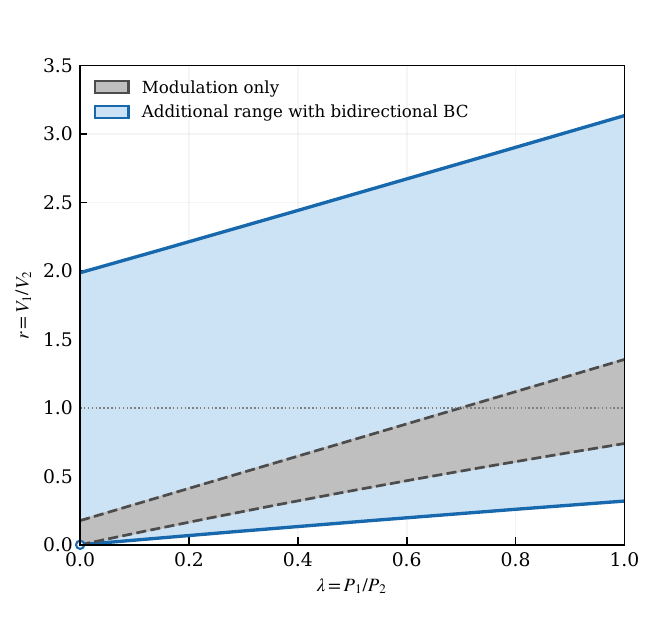}%
    }%
    \label{fig:bc_fourth}}

\end{minipage}
};

\node at ([xshift=-2mm,yshift=-19mm]main.center) {%
    \includegraphics[
        width=0.13\textwidth
    ]{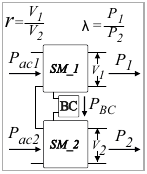}
    };
    
\node at ([xshift=-3 mm,yshift=76mm]main.center) {%
    \includegraphics[
        width=0.28\textwidth
    ]{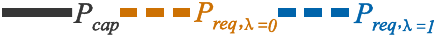}
    };

\end{tikzpicture}

\caption{BC power-transfer capability and achievable voltage regulation:
(a) forward-transfer capability and required balancing power for $\lambda=0$ and $1$ as functions of $r$ and $M_1$;
(b) the fixed-modulation section $M_1=M_2=M_0$;
(c) feasible forward-transfer combinations of $r$ and $M_1$;
(d) achievable voltage-ratio ranges versus load ratio, using modulation alone or joint modulation and BC operation.
Parameters: $M_0=0.85$, $\alpha=1$, $\ell=0.001$, and $\phi=1/6$.}
\label{fig:bc_analysis}

\end{figure*}

The calculation uses $M_0=0.85$, $\phi=1/6$, $\alpha=1$, and $\ell=0.001$. The carrier displacement is retained from the cascaded arrangement. With $\alpha=1$, module~2 operates at rated load, while
$\lambda$ specifies the load of module~1. The cases $\lambda=0$ and $\lambda=1$ represent an unloaded module~1 and equal rated loads, respectively. Fig.~\ref{fig:bc_analysis} illustrates how the BC power capability and modulation allocation determine the achievable voltage range.

\begin{table}[!t]
\centering
\caption{Calculated voltage ranges with adjustable modulation amplitudes}
\label{tab:vr_variable_modulation}
\renewcommand{\arraystretch}{1.15}
\setlength{\tabcolsep}{3pt}
\begin{tabular}{c l c c}
\hline
$\lambda$ & Method & $r=V_1/V_2$ & $V_2/V_{\mathrm{sm}}$\\
\hline
0 & Modulation only & $(0,\,0.176]$ & $[1.700,\,2)$\\
0 & Modulation + BC & $(0,\,1.986]$ & $[0.670,\,2)$\\
1 & Modulation only & $[0.739,\,1.353]$ & $[0.850,\,1.150]$\\
1 & Modulation + BC & $[0.319,\,3.133]$ & $[0.484,\,1.516]$\\
\hline
\end{tabular}
\end{table}

Fig.~\ref{fig:bc_analysis}(a) compares the forward BC capability with the balancing-power requirements for the two load cases. For each $(r,M_1)$, the second modulation amplitude is fixed by $M_2=M_0(1+r)-rM_1$, and the power allocation is $x=M_1r/[M_0(1+r)]$. The capability surface is common to both load cases, whereas the required power depends on $\lambda$. Only nonnegative demand is shown. Forward transfer is feasible wherever this demand does not exceed the capability at the same $(r,M_1)$.

Fig.~\ref{fig:bc_analysis}(b) shows the section with $M_1=M_2=M_0$. The capability--demand intersections give upper voltage-ratio bounds of approximately $1.870$ for $\lambda=0$ and $2.872$ for $\lambda=1$. These values provide the reference for assessing the benefit of adjustable modulation amplitudes.

Fig.~\ref{fig:bc_analysis}(c) shows the feasible forward-transfer combinations in the $r$--$M_1$ plane. Each shaded region satisfies both the modulation limits and $0\leq p_{\mathrm{req}}\leq p_{\mathrm{cap}}$ for the corresponding load case. The zero-demand boundary represents operation with the BC disabled. At a given $r$, varying $M_1$ changes both the required balancing power and the available transfer windows. This additional freedom extends the feasible voltage ratio beyond the fixed-modulation intersections in Fig.~\ref{fig:bc_analysis}(b).

Fig.~\ref{fig:bc_analysis}(d) gives the resulting voltage-ratio ranges under the assumed bidirectional symmetry. The gray band represents modulation-only operation, and the blue regions show the additional range enabled by the BC. The endpoint load cases are summarized in Table~\ref{tab:vr_variable_modulation}.

For $\lambda=0$, modulation-only operation requires $M_1=0$. The condition $M_2\leq1$ then restricts $V_2/V_{\mathrm{sm}}$ to $[1.7,2)$. Joint modulation and BC operation lowers the attainable input voltage to approximately $0.670V_{\mathrm{sm}}$, corresponding to $r_{\max}\simeq1.986$. At this boundary, module~1 receives approximately $60.6\%$ of the pair's ac input power, compared with $66.5\%$ under equal modulation amplitudes at the same voltage ratio. The smaller allocation reduces the power that must be transferred through the BC. The upper limit $V_2/V_{\mathrm{sm}}=2$ is excluded because it would require $V_1=0$.

For $\lambda=1$, joint operation expands the voltage interval from $0.85\leq V_2/V_{\mathrm{sm}}\leq1.15$ to approximately $0.484\leq V_2/V_{\mathrm{sm}}\leq1.516$. Under the assumed bidirectional symmetry, the corresponding voltage-ratio bounds are reciprocal: $r_{\min}\simeq0.319$ and $r_{\max}\simeq3.133$. The lower boundary involves reverse BC transfer, which is included in Fig.~\ref{fig:bc_analysis}(d).

At the upper joint boundaries, $M_2=1$, while $M_1\simeq0.774$ for $\lambda=0$ and $0.802$ for $\lambda=1$. The BC rms-current limit is reached in both cases; the normalized peak currents remain below their limit. Thus, the voltage boundaries are set jointly by the modulation and BC current constraints. These averaged-current results neglect switching ripple. The $\lambda=0$ voltage range is used in the following DAB comparison.

\section{Benefits for the Isolated DC--DC Stage}
\label{sec:dab_benefits}

The voltage range established in Section~\ref{sec:voltage_regulation_range} allows the isolated dc--dc converter input to follow its output-voltage requirement. This section compares a dual-active-bridge converter operating with a fixed input voltage and with voltage following enabled by joint CHB modulation and BC operation. Both cases use single-phase-shift (SPS) modulation and identical hardware and output requirements.

\subsection{Effect of Input-Voltage Adjustment}
\label{subsec:dab_voltage_effect}

The DAB inductor-current slope is set by the difference between the primary bridge voltage $u_{\mathrm P}$ and the primary-referred secondary voltage $u'_{\mathrm S}$: $L_D\,di_L/dt=u_{\mathrm P}-u'_{\mathrm S}$, where $L_D$ is the primary-referred series inductance and positive current flows toward the transformer. Unequal bridge-voltage amplitudes therefore cause the current to change even when the bridge voltages have the same polarity.

\begin{figure}[!t]
\centering
\subfloat[Fixed input voltage.\label{fig:dab_waveform_fixed}]{\includegraphics[width=\columnwidth]{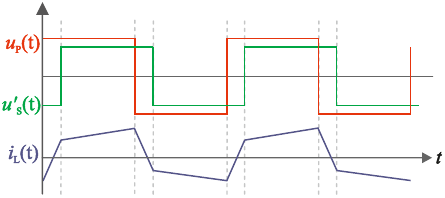}}

\vspace{2mm}
\subfloat[Voltage following.\label{fig:dab_waveform_following}]{\includegraphics[width=\columnwidth]{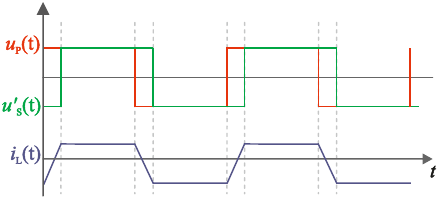}}
\caption{Schematic DAB waveforms at the same output voltage and power: (a) fixed input with $V_{\mathrm{in}}>n_TV_{\mathrm{out}}$ and (b) voltage following with matched bridge-voltage amplitudes. The secondary voltage is referred to the primary side.}
\label{fig:dab_waveform_comparison}
\end{figure}

Fig.~\ref{fig:dab_waveform_comparison} compares the waveforms at the same output voltage and power. With fixed input and $V_{\mathrm{in}}>n_TV_{\mathrm{out}}$, the current continues to rise when both bridge voltages are positive and falls when both are negative, where $n_T=N_{\mathrm p}/N_{\mathrm s}$. Reducing the input to $V_{\mathrm{in}}=n_TV_{\mathrm{out}}$ produces current plateaus during these intervals. The phase shift is adjusted to maintain the transferred power. This input-voltage adjustment changes both the current peaks and the commutation currents, providing the basis for reducing current stress and extending ZVS operation.

\subsection{Voltage Selection and Peak-Current Comparison}
\label{subsec:dab_basis}
\label{subsec:dab_voltage_selection}
\label{subsec:dab_currents}

An extreme operating condition is considered in this section, where module~2 operates at full load while module~1 is unloaded ($\lambda=0$),  so $V_{\mathrm{in}}=V_2$ and $P_{\mathrm{out}}=P_2$ when losses are neglected. The analysis focuses on the DAB conversion stage connected to module~2. For a fair comparison, the circuit parameters, switching frequency and transformer design are kept identical in the two operating schemes.

Using the nominal submodule voltage $V_{\mathrm{sm}}$ and rated power $P_{\mathrm{sm}}$ as fixed bases, define
\begin{equation}
G=\frac{n_TV_{\mathrm{out}}}{V_{\mathrm{sm}}},\qquad
k=\frac{V_{\mathrm{in}}}{V_{\mathrm{sm}}},\qquad
q=\frac{P_{\mathrm{out}}}{P_{\mathrm{sm}}}.
\label{eq:dab_normalization}
\end{equation}
Thus, $G$ specifies the output voltage, $k$ specifies the adjustable input voltage, and $q$ specifies the load. Both voltages share the same fixed base, and $k=G$ gives matched bridge-voltage amplitudes. For conventional wide-output-range applications, the output range is $0.4\leq G\leq2.5$, with the power limit
\begin{equation}
q_{\mathrm{lim}}(G)=\min\{G,1\}.
\label{eq:dab_power_envelope}
\end{equation}
This envelope gives constant output-current capability below nominal voltage and rated power above it. Peak current is compared along $q=q_{\mathrm{lim}}$, while ZVS is evaluated throughout $0<q\leq q_{\mathrm{lim}}$.

For $\lambda=0$, the joint-operation range in Section~\ref{subsec:vr_results} gives $0.670\lesssim k<2$. The numerical comparison uses an upper limit of $1.98$ to retain a positive module~1 voltage. This rated-load range is retained at partial load because the required balancing power decreases with load for a given modulation allocation. The two input commands are
\begin{equation}
k=
\begin{cases}
1, & \text{fixed input},\\
\min\{1.98,\max\{G,0.670\}\}, & \text{voltage following}.
\end{cases}
\label{eq:dab_following_piecewise}
\end{equation}
Voltage following therefore matches the bridge amplitudes over approximately $0.670\leq G\leq1.98$ and holds the input at its corresponding limit outside this interval. 
Since this extreme operating condition cannot be achieved by adjusting the modulation index alone, and the focus of this work is on the balancing circuit, the fixed-input baseline uses $V_1=V_2=V_{\mathrm{sm}}$ and $M_1=M_2=0.85$, with the BC providing the balancing power required by the unequal module loads.

For each operating point, the secondary bridge phase lag is adjusted to deliver the same output power. Let $\delta\in[0,1/2]$ denote this lag divided by $\pi$, and let $\kappa=8f_DL_DP_{\mathrm{sm}}/V_{\mathrm{sm}}^2$ describe the common DAB hardware. The SPS power relation and required phase shift are
\begin{equation}
\begin{aligned}
q&=\frac{4kG}{\kappa}\delta(1-\delta),\\
\delta&=\frac{1-\sqrt{1-\kappa q/(kG)}}{2}.
\end{aligned}
\label{eq:dab_required_phase}
\end{equation}
A $30^\circ$ lag at the nominal matched rated point, $k=G=q=1$, gives $\kappa=5/9$. Both input commands can supply the specified output envelope.

The current extrema follow from the two bridge transitions. Taking the primary positive transition as $t=0$, the secondary transition occurs at $t=\delta/(2f_D)$. Let $j_0$ and $j_1$ denote the currents at these transitions normalized by $I_b=P_{\mathrm{sm}}/V_{\mathrm{sm}}$. Half-wave symmetry gives
\begin{equation}
\begin{aligned}
j_0&=\frac{2}{\kappa}\big[-k+G(1-2\delta)\big],\\
j_1&=\frac{2}{\kappa}\big[-k(1-2\delta)+G\big],\\
\frac{I_{L,\mathrm{pk}}}{I_b}&=\max\{|j_0|,|j_1|\}.
\end{aligned}
\label{eq:dab_currents}
\end{equation}
The peak follows from the piecewise-linear half-cycle trajectory $j_0\rightarrow j_1\rightarrow-j_0$.

\begin{figure}[!t]
\centering
\includegraphics[width=\columnwidth]{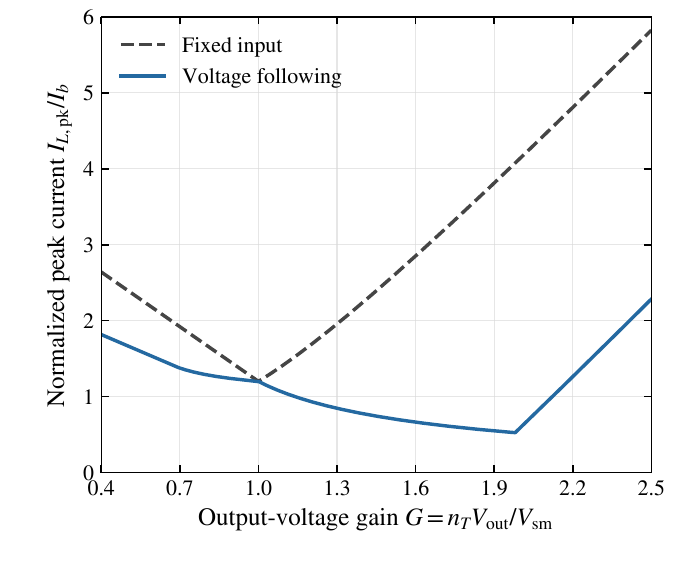}
\caption{Normalized DAB peak current with fixed input and voltage following along $q=\min\{G,1\}$. Voltage following uses joint modulation and BC operation with $0.670\lesssim k\leq1.98$. Both cases use $\lambda=0$ and $\kappa=5/9$.}
\label{fig:dab_peak_comparison}
\end{figure}

Fig.~\ref{fig:dab_peak_comparison} shows the resulting peak-current reduction. At $G=0.4$, voltage following reduces the peak from $2.64I_b$ to $1.82I_b$, a reduction of $31.2\%$. At $G=1$, both cases give $1.20I_b$ because their input voltages coincide. At $G=2$, the following input remains close to the referred output voltage, reducing the peak from $4.14I_b$ to $0.591I_b$ ($85.7\%$). Above the matching interval, the increasing voltage mismatch causes the following current to rise again. At $G=2.5$, it reaches $2.28I_b$, compared with $5.83I_b$ for fixed input.

\subsection{Extension of the Ideal ZVS Region}
\label{subsec:dab_zvs}

With the stated current direction, ideal ZVS in both bridges requires $j_0<0$ and $j_1>0$. Combining \eqref{eq:dab_currents} and \eqref{eq:dab_required_phase} gives the minimum-load boundary
\begin{equation}
q_Z(k,G)=\frac{kG}{\kappa}
\left[1-\min\left\{\frac{k}{G},\frac{G}{k}\right\}^{2}\right].
\label{eq:dab_zvs_power}
\end{equation}
The two-bridge ZVS condition is $q>q_Z$; equality corresponds to zero current at a commutation. This ideal boundary uses current polarity and neglects device capacitance and dead time.

For fixed input, substituting $k=1$ gives a zero minimum-load boundary only at $G=1$. Voltage following instead gives $q_Z=0$ throughout its matching interval, approximately $0.670\leq G\leq1.98$. A positive minimum-load requirement returns when the input reaches either limit.

\begin{figure}[!t]
\centering
\includegraphics[width=\columnwidth]{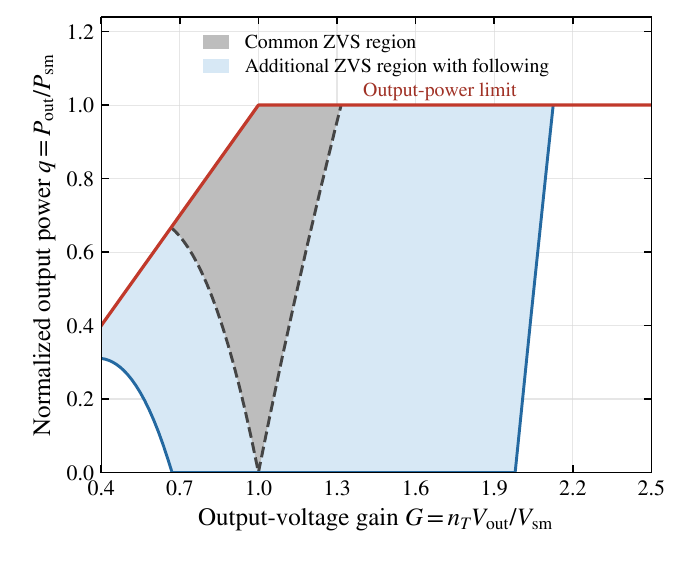}
\caption{Ideal two-bridge ZVS regions with fixed input and voltage following for $\lambda=0$ and $\kappa=5/9$. Gray denotes the common ZVS region; blue denotes the additional region enabled by voltage following. The red curve is the output-power limit $q_{\mathrm{lim}}=\min\{G,1\}$.}
\label{fig:dab_zvs_region}
\end{figure}

Fig.~\ref{fig:dab_zvs_region} shows the common ZVS region in gray and the additional region obtained through voltage following in blue. Unshaded points below the red output-power limit do not satisfy simultaneous two-bridge ZVS under either input command.

Along $q=q_{\mathrm{lim}}$, the ideal ZVS interval expands from approximately $0.667<G<1.316$ with fixed input to $0.4\leq G<2.125$ with voltage following. The latter extends beyond the upper matching limit of $G=1.98$: after input saturation, the remaining voltage mismatch is still small enough to maintain ZVS until the minimum-load boundary reaches the output-power limit.

\section{Experimental Verification}
\label{sec:experiment}

A six-submodule CHB-SST prototype was used to experimentally verify the proposed variable dc-link voltage operation. Embedded BCs are connected between adjacent submodules. In the experiments presented below, $SM_1$--$SM_3$ are used to demonstrate intentionally unequal dc-link voltage operation through the four-switch buck--boost function, whereas the remaining submodules are operated with equal dc-link voltages under conventional balancing operation. 

\begin{figure}[!t]
    \centering
    \includegraphics[width=\columnwidth]{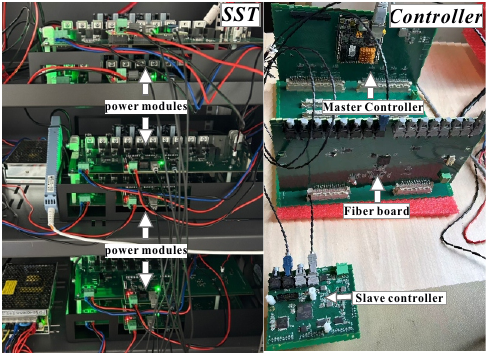}
    \caption{Six-submodule SST experimental prototype.}
    \label{fig:proto}
\end{figure}

The experimental prototype is shown in Fig.~\ref{fig:proto}, and the corresponding parameters are listed in Table~\ref{Table1}. The setup mainly consists of a ZYNQ-7020 FPGA-based central control platform and silicon-MOSFET-based power modules, with each submodule equipped with a dedicated slave controller implemented on an XC7S15 FPGA.

\begin{table}[!t]
\centering
\renewcommand{\arraystretch}{1.5} 
\caption{Prototype Parameters}
\label{Table1}
\begin{tabular}{lll}
\hline
Parameter                     & Symbol & Value \\ \hline
AC Voltage                    & $U_\text{AC}$             & 250 V  \\
Number of submodules          & $N$                       & 6 \\
Grid frequency                & $f_\text{grid}$           & 50 Hz\\
AC Switching frequency           & $f_\text{CHB}$            & 5 kHz    \\
DC/DC Switching frequency           & $f_\text{DAB}$            & 100 kHz    \\
BC switching frequency        & $f_\text{BC}$             & 100 kHz    \\
Inductance of bridge inductor & $L_\text{AC}$             & 5 mH    \\
Turns ratio of HFT            & $n_\text{1}:n_\text{2}$   & 1:1    \\
Submodule DC capacitor        & $C_\text{DC}$             & $3000\ \mu\text{F}$  \\
Balancing inductor            & $L_\text{B}$              & $10\ \mu\text{H}$   \\ \hline
\end{tabular}
\end{table}

\begin{figure}[!t]
    \centering
    \includegraphics[width=\columnwidth]{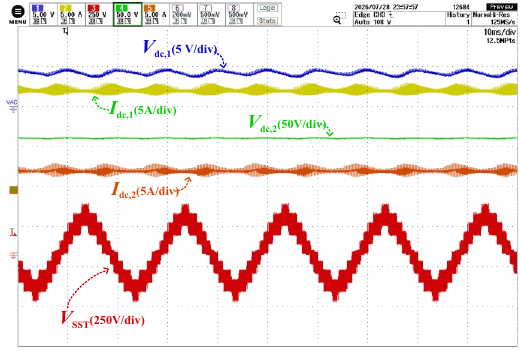}
\caption{Measured steady-state waveforms of the six-submodule CHB-SST under unequal dc output conditions, showing the output voltages $V_{\mathrm{dc},1}$ and $V_{\mathrm{dc},2}$, output currents $I_{\mathrm{dc},1}$ and $I_{\mathrm{dc},2}$, and the SST ac-side voltage $V_{\mathrm{sst}}$.}
    \label{fig:exp_sst}
\end{figure}

Fig.~\ref{fig:exp_sst} shows representative steady-state waveforms of the complete six-submodule CHB-SST. The two measured dc output ports operate at substantially different voltage levels, while both output currents are the same. Meanwhile, the SST voltage exhibits the expected multilevel waveform. The result demonstrates that the complete cascaded converter maintains normal steady-state operation while providing dc ports with significantly different voltage and power requirements.

\begin{figure}[!t]
    \centering
    \includegraphics[width=\columnwidth]{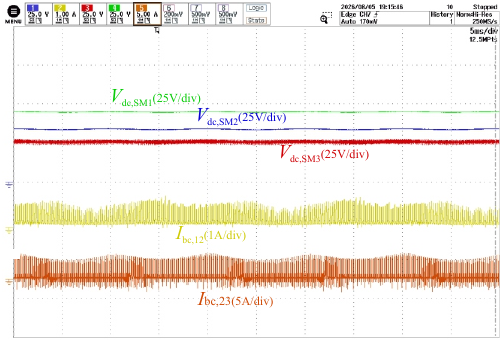}
    \caption{Measured dc-link voltages and BC currents with $V_{\mathrm{dc},1}=45$~V, $V_{\mathrm{dc},2}=35$~V, and $V_{\mathrm{dc},3}=25$~V.}
    \label{fig:exp_voltage_1}
\end{figure}

\begin{figure}[!t]
    \centering
    \includegraphics[width=\columnwidth]{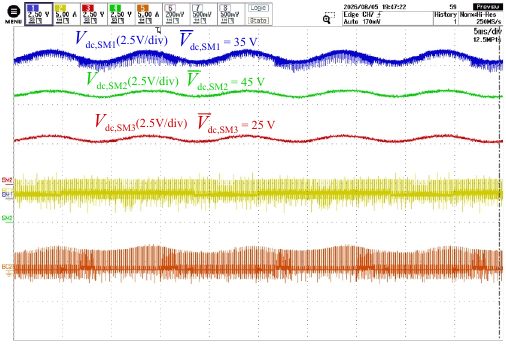}
    \caption{Measured dc-link voltages and BC currents with $V_{\mathrm{dc},1}=35$~V, $V_{\mathrm{dc},2}=45$~V, and $V_{\mathrm{dc},3}=25$~V.}
    \label{fig:exp_voltage_2}
\end{figure}

The capability to establish intentionally different submodule dc-link voltage levels is further demonstrated in Figs.~\ref{fig:exp_voltage_1} and~\ref{fig:exp_voltage_2}. In Fig.~\ref{fig:exp_voltage_1}, the measured dc-link voltages of $SM_1$, $SM_2$, and $SM_3$ are approximately 45~V, 35~V, and 25~V, respectively. Thus, the adjacent dc links operate with $V_{\mathrm{dc},1}>V_{\mathrm{dc},2}>V_{\mathrm{dc},3}$ rather than converging toward a common voltage.

A different operating point is shown in Fig.~\ref{fig:exp_voltage_2}, where the voltage ordering of the first two submodules is reversed. In this case, $V_{\mathrm{dc},1}$, $V_{\mathrm{dc},2}$, and $V_{\mathrm{dc},3}$ are approximately 35~V, 45~V, and 25~V, respectively. Comparison of Figs.~\ref{fig:exp_voltage_1} and~\ref{fig:exp_voltage_2} shows that the relative dc-link voltage levels are not inherently fixed by the cascaded structure or by the balancing network. Instead, different unequal-voltage operating points can be established through the active four-switch BC operation. In particular, both $V_{\mathrm{dc},1}>V_{\mathrm{dc},2}$ and $V_{\mathrm{dc},1}<V_{\mathrm{dc},2}$ are experimentally demonstrated. This behavior is consistent with the buck--boost operating principle analyzed in Section~\ref{subsec:buck_boost_states} and distinguishes the proposed operation from conventional balancing schemes.

\begin{figure}[!t]
    \centering
    \includegraphics[width=\columnwidth]{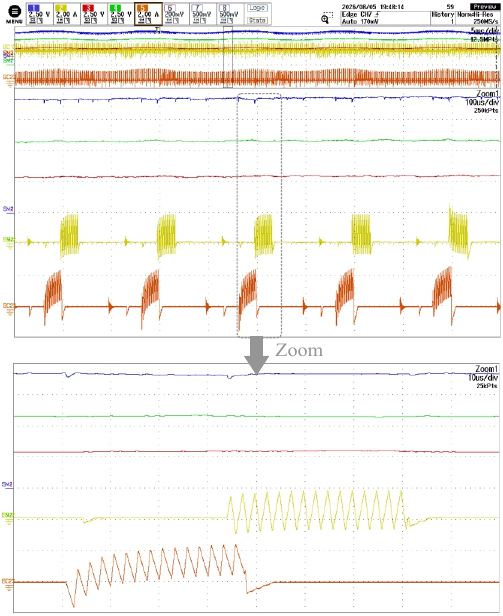}
    \caption{Measured balancing-inductor current waveforms of the embedded BCs under unequal dc-link voltage operation.}
    \label{fig:exp_bc_current}
\end{figure}

The detailed balancing-current waveforms are shown in Fig.~\ref{fig:exp_bc_current}. The measured currents of the BCs between the three unequal-voltage submodules exhibit repeated high-frequency current-transfer intervals. These waveforms show repeated BC-current conduction while the adjacent dc-link voltages remain intentionally different.

\begin{figure}[!t]
\centering
\subfloat[\(V_{\mathrm{in}}=n_TV_{\mathrm{out}}\),\label{fig:dab1}]{\includegraphics[width=\columnwidth]{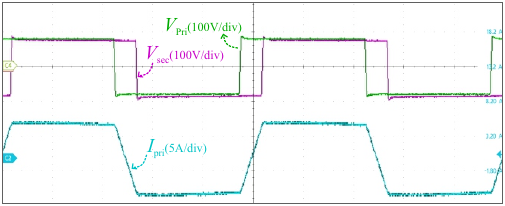}}

\vspace{0mm}
\subfloat[\(V_{\mathrm{in}}>n_TV_{\mathrm{out}}\),\label{fig:dab2}]{\includegraphics[width=\columnwidth]{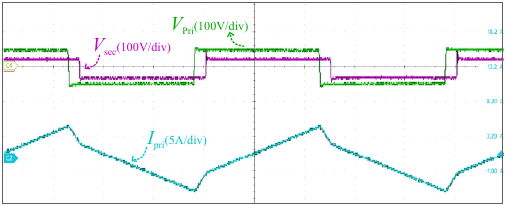}}

\subfloat[\(V_{\mathrm{in}}<n_TV_{\mathrm{out}}\)\label{fig:dab3}]{\includegraphics[width=\columnwidth]{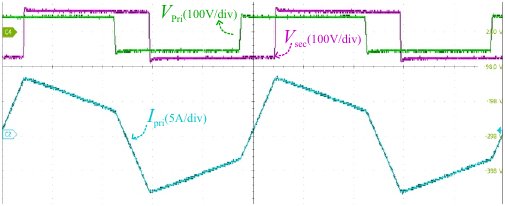}}

\caption{Measured DAB waveforms showing the primary-side voltage \(V_{\mathrm{pri}}\), secondary-side voltage \(V_{\mathrm{sec}}\), and primary current \(I_{\mathrm{pri}}\) under different voltage-conversion conditions.}
\label{fig:dab}
\end{figure}

\begin{figure}[!t]
    \centering
    \includegraphics[width=\columnwidth]{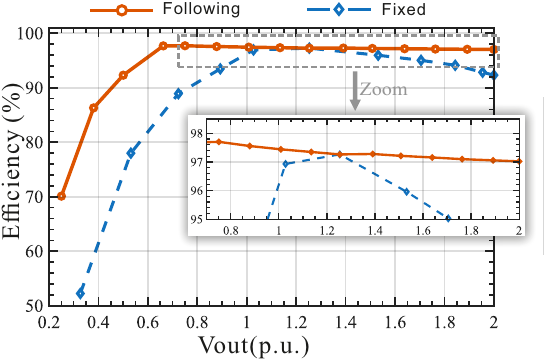}
    \caption{Measured DAB efficiency versus normalized output voltage under fixed input-voltage operation and output-following input-voltage operation.}
    \label{fig:efficiency}
\end{figure}

Fig.~\ref{fig:dab} presents the measured transformer voltage and current waveforms of the DAB converter under three voltage-conversion conditions, namely \(V_{\mathrm{in}}>n_TV_{\mathrm{out}}\), \(V_{\mathrm{in}}=n_TV_{\mathrm{out}}\), and \(V_{\mathrm{in}}<n_TV_{\mathrm{out}}\). The measured waveforms show that voltage matching suppresses the current variation during the same-polarity bridge-voltage intervals, consistent with the reduced circulating-current behavior predicted in Section IV. This result is consistent with the analysis in Section~\ref{sec:dab_benefits}, where voltage matching reduces the circulating current in the isolated conversion stage.

Fig.~\ref{fig:efficiency} compares the measured DAB efficiency under fixed-input and voltage-following operation over output-voltage. The voltage base is \(V_{\rm base}=50\) V. Both tests use the same converter hardware, switching frequency, and resistive load, ensuring the same output power at each compared output voltage. In the fixed-input case, the input voltage remains constant. In the voltage-following case, it tracks the output voltage within an approximate range of 0.67--2.0~p.u., consistent with the range obtained in Section~III for $\lambda=0$. Voltage following provides the largest efficiency improvement at low output voltages, where it reduces the substantial voltage mismatch associated with fixed-input operation.
The efficiency difference becomes small near the matching point of the fixed-input case, while voltage following retains an advantage at higher output voltages.
These results demonstrate the efficiency benefit of adjusting the DAB input voltage within a finite available range.

Overall, the experimental results demonstrate the proposed concept at both the system and submodule levels. The complete six-submodule CHB-SST maintains normal multilevel operation while supplying dc outputs with significantly different voltage and power requirements. Within the cascaded converter, intentionally unequal submodule dc-link voltages are experimentally established at different voltage orderings, while the embedded BCs provide the required inter-submodule energy-transfer paths. These results establish the functionality of unequal-voltage operation. The benefits of adjusting the submodule dc-link voltage are also experimentally verified in the DAB stage, including a clear reduction in transformer current and an overall improvement in conversion efficiency.

\section{Conclusion}
\label{sec:conclusion}

This paper investigated variable dc-link voltage operation of a single-isolation-stage multioutput SST through coordinated CHB modulation and embedded balancing circuits. By relating the required
balancing power to the BC transfer capability under current constraints, the achievable submodule voltage range was established. The analysis showed that BC power exchange extends the range
available through modulation alone, enabling DAB input-voltage adjustment to reduce peak current and extend the ideal ZVS region.
Experiments on a six-submodule SST prototype demonstrated unequal dc-link voltage operation and inter-submodule power exchange. Separate DAB tests demonstrated reduced current stress and improved efficiency with voltage following compared with fixed input. These results support the use of adjustable submodule dc-link voltages to accommodate wide-range dc outputs.

\bibliographystyle{IEEEtran} 
\bibliography{ref}

\vfill

\end{document}